\documentclass[iop]{emulateapj}

\newcommand{\helium}{\ion{He}{1}~$\lambda$10830}

\shorttitle{Outbursting Protostar in L 1641}
\shortauthors{Fischer et al.}

\begin{document}

\title{Multiwavelength Observations of V2775 Ori, an Outbursting Protostar in L 1641:\\Exploring the Edge of the FU Orionis Regime}
\author{William J. Fischer\altaffilmark{1}, S. Thomas Megeath\altaffilmark{1}, John J. Tobin\altaffilmark{2}, Amelia M. Stutz\altaffilmark{3,4}, Babar Ali\altaffilmark{5}, Ian Remming\altaffilmark{6}, Marina Kounkel\altaffilmark{1}, Thomas Stanke\altaffilmark{7}, Mayra Osorio\altaffilmark{8}, Thomas Henning\altaffilmark{3}, P. Manoj\altaffilmark{6}, and T. L. Wilson\altaffilmark{9}}
\altaffiltext{1}{Department of Physics and Astronomy, University of Toledo, 2801 W.\ Bancroft St., Toledo, OH 43606, USA; wfische@utnet.utoledo.edu}
\altaffiltext{2}{National Radio Astronomy Observatory, 520 Edgemont Rd., Charlottesville, VA 22903, USA}
\altaffiltext{3}{Max-Planck-Institut f\"ur Astronomie, K\"onigstuhl 17, D-69117 Heidelberg, Germany}
\altaffiltext{4}{Steward Observatory, University of Arizona, 933 North Cherry Avenue, Tucson, AZ 85721, USA} 
\altaffiltext{5}{NHSC/IPAC/Caltech, 770 S. Wilson Avenue, Pasadena, CA 91125, USA}
\altaffiltext{6}{Department of Physics and Astronomy, 500 Wilson Blvd., University of Rochester, Rochester, NY 14627, USA}
\altaffiltext{7}{ESO, Karl-Schwarzschild-Strasse 2, 85748 Garching bei M\"unchen, Germany}
\altaffiltext{8}{Instituto de Astrof\'{i}sica de Andaluc\'{i}a, CSIC, Camino Bajo de Hu\'{e}tor 50, E-18008, Granada, Spain}
\altaffiltext{9}{Naval Research Laboratory, 4555 Overlook Ave.\ SW, Washington, DC 20375, USA}

\begin{abstract}
Individual outbursting young stars are important laboratories for studying the physics of episodic accretion and the extent to which this phenomenon can explain the luminosity distribution of protostars.  We present new and archival data for V2775 Ori (HOPS 223), a protostar in the L 1641 region of the Orion molecular clouds that was discovered by Caratti o Garatti et al.\ (2011) to have recently undergone an order-of-magnitude increase in luminosity.  Our near-infrared spectra of the source have strong blueshifted \helium\ absorption, strong H$_2$O and CO absorption, and no \ion{H}{1} emission, all typical of FU Orionis sources.  With data from IRTF, 2MASS, HST, Spitzer, WISE, Herschel, and APEX that span from 1 to 70 \micron\ pre-outburst and from 1 to 870 \micron\ post-outburst, we estimate that the outburst began between 2005 April and 2007 March.  We also model the pre- and post-outburst spectral energy distributions of the source, finding it to be in the late stages of accreting its envelope with a disk-to-star accretion rate that increased from $\sim2\times10^{-6}~M_\sun~\rm{yr}^{-1}$  to $\sim10^{-5}~M_\sun~\rm{yr}^{-1}$ during the outburst.  The post-outburst luminosity at the epoch of the FU Orionis-like near-IR spectra is $28~L_\sun$, making V2775 Ori the least luminous documented FU Orionis outburster with a protostellar envelope.  The existence of low-luminosity outbursts supports the notion that a range of episiodic accretion phenomena can partially explain the observed spread in protostellar luminosities.
\end{abstract}


\section{INTRODUCTION}\label{s.intro}

Stochastic luminosity outbursts are the most extreme type of photometric variability observed in young stellar objects.  These outbursts range from the less dramatic EX Lupi-type sources, which occasionally climb in luminosity by a factor of a few, fade over a few years, and have the emission-line spectra of extreme T Tauri stars \citep{her07}, to the most dramatic FU Orionis-type sources, which increase in optical brightness by at least four magnitudes, remain luminous for decades, and have absorption-line spectra with inferred spectral types that change with wavelength \citep{har96}.  Based on their mid-infrared spectra, FU Orionis sources have been divided into Category 1, which have the 10~\micron\ silicate feature in absorption and still have infalling envelopes, and Category 2, which have silicate emission and lack significant envelopes \citep{qua07}.  This suggests that these outbursts are an important phenomenon at all stages in the evolution of young stellar objects.

The outbursts are due to luminosity generated by an increase in the rate at which the circumstellar disk material accretes onto the star.  The notion that the disk-to-star accretion rate of a young stellar object is typically low but occasionally spikes during times of dramatically enhanced accretion activity is known as episodic accretion.  This was proposed to explain why the observed luminosities of populations of protostars (young stellar objects with envelopes) have both systematically lower luminosities and a wider dispersion in luminosities than those predicted for the collapse of isothermal spheres \citep[the luminosity problem;][]{ken90,eva09,dun10}.  If each forming star experiences several relatively short but large luminosity outbursts, its time-averaged luminosity could agree with predictions even if the star is usually underluminous.

Outbursts have been modeled as the accretion of clumps formed via instabilities in a self-gravitating disk \citep{vor05,vor06,vor10,zhu09,zhu10}.  In these models, the infall of gas from the envelope causes the disk to grow in mass until it becomes gravitationally unstable.  An open question is whether these events are frequent enough to solve the luminosity problem.  \citet{gre08} distinguish between sources in which an outburst was actually observed and those that merely have outburst-like spectra, pointing to the importance of time monitoring to unambiguously classify each putative outburster and determine the frequency of these events.  Since the advent of large-scale optical and IR monitoring programs, several young stellar objects have been observed to undergo significant increases in luminosity; e.g., V733 Cep \citep{rei07}, V2492 Cyg \citep{cov11}, V2493 Cyg \citep{sem10,mil11}, V2494 Cyg \citep{asp09}, V2495 Cyg \citep{mov06}, V900 Mon \citep{rei12}, and V1647 Ori \citep{rei04,abr04,bri04}, allowing a more thorough understanding of the range of outburst types and their frequencies.

Caratti o Garatti et al.\ (2011, hereafter \citealt{car11}) announced one such outburst, in which a young stellar object in the Orion molecular clouds underwent an order-of-magnitude increase in its luminosity.  The source, near the southern edge of the L 1641 region at $\alpha=5^h42^m48^s.48$, $\delta=-8^\circ16'34''.7$ (J2000), has the designation 2MASS J05424848-0816347.  The discoverers of the outburst named the source [CTF93] 216-2 due to its proximity to source 216 of \citet{che93}.

The source appears in the catalog for HOPS \citep{fis10,sta10,ali10}, the Herschel Orion Protostar Survey, an open-time key program of the Herschel Space Observatory.  The goal of HOPS is to obtain 70 and 160~\micron\ imaging and 50--200~\micron\ spectra of protostars in the Orion molecular clouds that were identified with Spitzer Space Telescope photometry by \citet{meg12}.  HOPS is the centerpiece of an effort to acquire images, photometry, and spectra from near-infrared to millimeter wavelengths of these targets, yielding the most complete collection of data on the largest star-forming region within 500 pc of the Sun.  In the HOPS catalog, the source is designated HOPS 223.

Throughout this paper we use the name V2775 Ori.  This is the source's recently announced designation in the General Catalogue of Variable Stars \citep{kaz11}.  We assume a distance to the source $d=420~{\rm pc}$ from recent high-precision parallax studies of non-thermal sources in the Orion Nebula region \citep{san07,men07,kim08}. 

In this paper we present archival near-to-mid-IR photometry and spectra for V2775 Ori as well as a new near-IR image, new far-IR and submillimeter imaging and photometry, and new near-IR moderate-resolution spectra.  We present a new estimate for the timing of the outburst and model the evolving spectral energy distribution (SED) of the source to constrain its envelope density and luminosity.  With the near-IR spectra, we present the first published \helium\ profiles of the source, an important accretion and outflow diagnostic \citep{edw06}.  We do not detect atomic hydrogen emission lines, which are strong in the spectra of heavily accreting young stars but absent from FU Orionis-type spectra \citep{con10}.

Section 2 of the paper describes new and archival images, photometry, and spectra of the source, Section 3 analyzes the new images of the source and its environment, Section 4 discusses the timing of the outburst, Section 5 uses the SED to constrain the properties of the source, and Section 6 uses the source's luminosity and near-IR spectrum to analyze the nature of the outburst.  Conclusions are in Section 7.

\section{OBSERVATIONS}

\begin{figure*}
\plotone{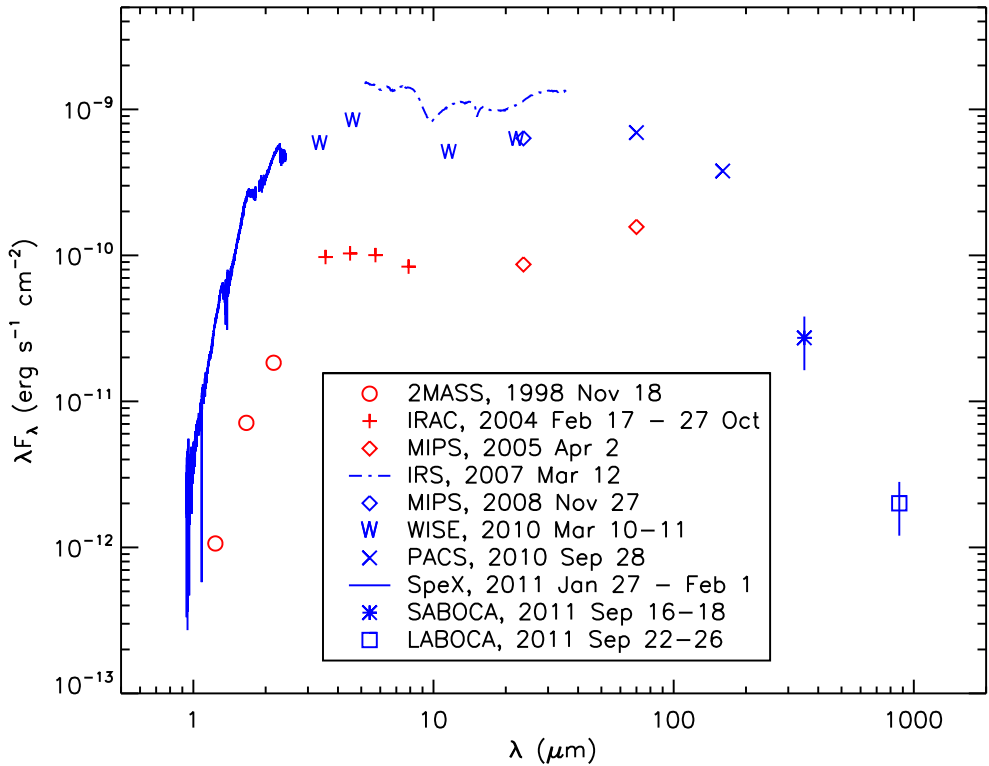}
\figcaption{Photometry and spectra of V2775 Ori.  Pre-outburst data are in red, and post-outburst data are in blue, placing the outburst between 2005 April 2 and 2007 March 12.  Photometric uncertainties are shown only at 350 and 870 \micron; elsewhere they are smaller than the symbols.\label{f.sed}}
\end{figure*}

We report new and archival imaging, photometry, and spectra for V2775 Ori, extending from 1.0 to 870 \micron\ and covering nearly a thirteen-year interval that contains the outburst event.  Photometry and uncertainties are reported in Tables~\ref{t.photometry1} and \ref{t.photometry2}, and the combined photometry and spectra are displayed in Figure~\ref{f.sed}.  We treat the object as a point source due to the lack of evidence for binarity and the positional coincidence across the wavelength range in imaging with angular resolutions extending from 0.2\arcsec\ (80 AU) to 19\arcsec.

\begin{deluxetable}{ccccc}
\tablecaption{Pre-Outburst Photometry for V2775 Ori\label{t.photometry1}}
\tablewidth{\hsize}
\tablehead{\colhead{$\lambda$} & \colhead{$F_\nu$} & \colhead{$\Delta F_\nu$} & \colhead{Instrument} & \colhead{Date} \\ \colhead{(\micron)} & \colhead{(mJy)} & \colhead{(mJy)} & \colhead{} & \colhead{}}
\startdata
1.2 & 0.438 & 0.0456 & 2MASS & 1998 Nov 18 \\
1.7 & 3.95 & 0.0873 & 2MASS & 1998 Nov 18 \\
2.2 & 13.2 & 0.329 & 2MASS & 1998 Nov 18 \\
3.6  & 115 & 5.77 & IRAC & 2004 Feb 17--Oct 27 \\
4.5  & 155 & 7.76 & IRAC & 2004 Feb 17--Oct 27 \\
5.8  & 192 & 9.60 & IRAC & 2004 Feb 17--Oct 27 \\
8.0  & 220 & 11.0 & IRAC & 2004 Feb 17--Oct 27 \\
24   & 685 & 34.8 & MIPS & 2005 Apr 2 \\
70   & 3660 & 428 & MIPS & 2005 Apr 2
\enddata
\end{deluxetable}

\begin{deluxetable}{ccccc}
\tablecaption{Post-Outburst Photometry for V2775 Ori\label{t.photometry2}}
\tablewidth{\hsize}
\tablehead{\colhead{$\lambda$} & \colhead{$F_\nu$} & \colhead{$\Delta F_\nu$} & \colhead{Instrument} & \colhead{Date} \\ \colhead{(\micron)} & \colhead{(mJy)} & \colhead{(mJy)} & \colhead{} & \colhead{}}
\startdata
3.4  & 593 & 18.6 & WISE & 2010 Mar 10--11 \\
4.6  & 1170 & 29.0 & WISE & 2010 Mar 10--11 \\
12  & 1780 & 19.7 & WISE & 2010 Mar 10--11 \\
22  & 4170 & 26.9 & WISE & 2010 Mar 10--11 \\
24   & 5010 & 257 & MIPS & 2008 Nov 27 \\
70   & 16200 & 810 & PACS & 2010 Sep 28 \\
160 & 20200 & 1020 & PACS & 2010 Sep 28 \\
350 & 3180 & 1270 & SABOCA & 2011 Sep 16--18 \\
870 & 582 & 233 & LABOCA & 2011 Sep 22--26
\enddata
\end{deluxetable}

\subsection{HST Imaging}

\begin{figure}
\resizebox{\hsize}{!}{\includegraphics{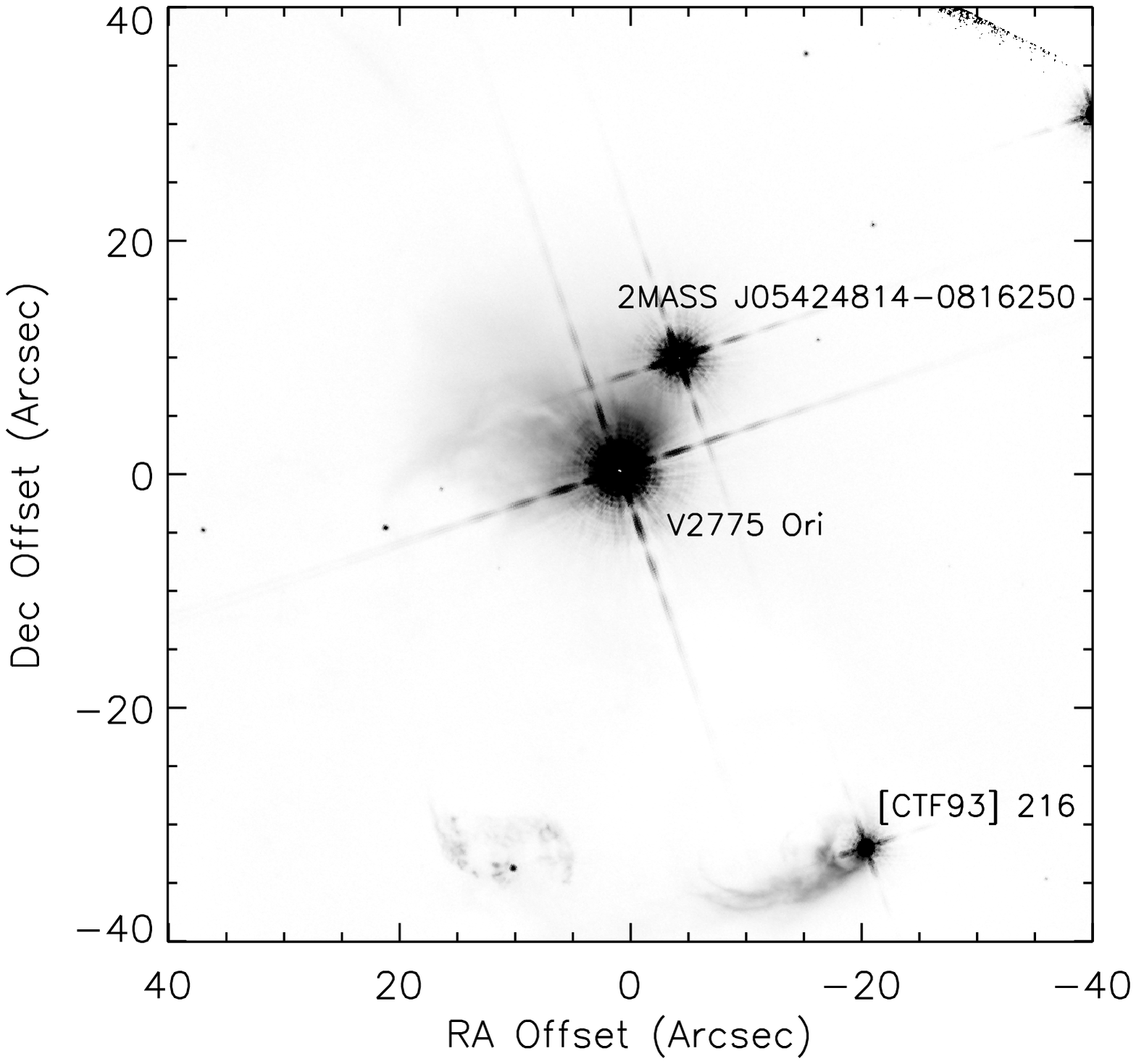}}
\figcaption{HST/WFC3 post-outburst 1.60 \micron\ image of the V2775 Ori region.  Offsets are from the Spitzer-determined J2000 position of V2775 Ori: $\alpha=5^h42^m48^s.48$, $\delta=-8^\circ16'34''.7$.  V2775 Ori has a reflection nebula typical of FU Orionis objects to its northeast, and nebulosity is also associated with the source 39\arcsec\ to the southwest ([CTF93] 216; 2MASS J05424706-0817069; HOPS 221).  The other point source, 11\arcsec\ to the northwest of V2775 Ori, is 2MASS J05424814-0816250.  This and subsequent images are shown on an inverse hyperbolic sine scale that reduces the intensity of the brightest regions to emphasize faint extended structure.\label{f.wfc3}}
\end{figure}

We used Wide Field Camera 3 (WFC3; \citealt{kimb08}) aboard HST to map a field containing V2775 Ori on 2010 April 14.  We used the IR detector and the F160W (1.60 \micron) filter.  The map was reduced using the instrument pipeline from a series of 5 dithers with total integration time 2496~s.  The final map is $145\arcsec \times 129\arcsec$, about 0.30 pc by 0.26 pc at the distance of Orion, and the angular resolution is 0.2\arcsec\ (0.065\arcsec/pixel), about 80~AU at the distance of Orion.  Sources brighter than 12 mag in the $H$ band, including HOPS 223, are saturated, but the faint extended emission associated with the source is cleanly imaged.  An 80\arcsec\ square cutout of the HST/WFC3 image appears in Figure~\ref{f.wfc3}. 

\subsection{2MASS, WISE, and Spitzer Photometry}

We obtained photometry at $J$, $H$, and $K_s$ from the Two Micron All Sky Survey (2MASS; \citealt{skr06}), which observed V2775 Ori on 1998 November 18.  We obtained photometry at 3.4, 4.6, 12, and 22~\micron\ from the Wide-field Infrared Survey Explorer (WISE; \citealt{wri10}), which observed the source on 2010 March 10--11.\footnote{Archival 2MASS and WISE photometry are available at http://irsa.ipac.caltech.edu/.}  The WISE photometry reported here is from the March 2012 All-Sky Release.

Photometry at 3.6, 4.5, 5.8, 8.0, and 24~\micron\ was obtained as part of a joint survey of the Orion A and B molecular clouds by the Infrared Array Camera (IRAC; \citealt{faz04}) and Multiband Imaging Photometer (MIPS; \citealt{rie04}) aboard Spitzer.  A detailed accounting of the Spitzer observations, data reduction, and source extraction can be found in \citet{kry12} and \citet{meg12}.  Here we summarize the most important details.

The IRAC observations of L 1641 were taken as part of Guaranteed Time Observation program 43 and were obtained in two epochs, one on 2004 February 18--19 and the other on 2004 October 8 and 27.  IRAC photometry was obtained at 3.6, 4.5, 5.8, and 8.0~\micron\ using an aperture of radius 2.4\arcsec\ with a sky annulus extending from 2.4\arcsec\ to 7.2\arcsec\ and corrected to infinity by dividing by aperture corrections:\ 0.824, 0.810, 0.725, and 0.631 in order of increasing wavelength.  Photometric zero points and zero-magnitude fluxes for the IRAC bands can be found in \citet{meg12} and \citet{rea05}, respectively.  The error estimate is dominated by calibration uncertainties, which we estimate to be 5\% in all channels.

The first MIPS observations of L 1641 were taken as part of Guaranteed Time Observation program 47 on 2005 April 2--3, and additional MIPS observations were taken as part of Guaranteed Time Observation program 50070 on 2008 November 27.  In both epochs, MIPS photometry was obtained at 24~\micron\ by fitting a point-spread function \citep{kry12}.  A 70~\micron\ image was available only at the first epoch.   At this wavelength, we performed aperture photometry with an aperture of radius 16\arcsec\ and a sky annulus extending from 18\arcsec\ to 39\arcsec.  We calibrated the 70~\micron\  photometry by comparing MIPS fluxes to Herschel/PACS fluxes (see the next subsection) at this wavelength for 72 sources across L 1641. At 24~\micron, the MIPS error estimates are dominated by a 5\% calibration uncertainty.  At 70~\micron, we estimate a 12\% uncertainty by adding in quadrature the calibration uncertainty of PACS (5\%) and the spread in the ratio of PACS to MIPS fluxes (11\%).

\subsection{Herschel/PACS Imaging and Photometry}

\begin{figure}
\resizebox{\hsize}{!}{\includegraphics{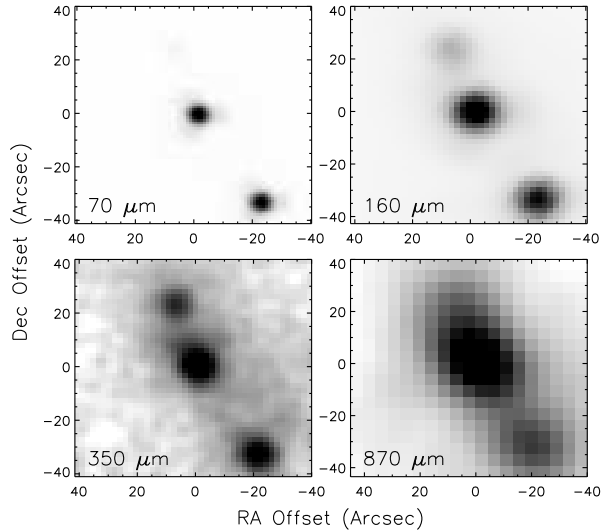}}
\figcaption{Post-outburst images of the V2775 Ori region at 70 and 160 \micron\ (Herschel/PACS), 350 \micron\ (APEX/SABOCA), and 870~\micron\ (APEX/LABOCA).  Offsets are from the Spitzer-determined position of V2775 Ori.  In addition to V2775 Ori (center), two additional sources are visible: [CTF93] 216 is in the southwest, and a newly identified point source is visible 25\arcsec\ to the northeast.  \label{f.farir}}
\end{figure}

Herschel observed an $8'$ square field containing V2775 Ori on 2010 September 28 (observing day 502; observation IDs 1342205254 and 1342205255) in the 70 $\mu$m and 160 $\mu$m bands available with the Photodetector Array Camera and Spectrometer (PACS; \citealt{pog10}), which have angular resolutions of 5.2\arcsec\ and 12\arcsec, respectively.  We observed our target field with homogeneous coverage using two orthogonal scanning directions and a scan speed of $20''$/s.  Each scan was repeated 6 times for a total observation time of 1229~s per scan direction.  Cutouts from the PACS fields containing V2775 Ori appear in the top row of Figure~\ref{f.farir}; these were reduced with the Scanamorphos package\footnote{http://www2.iap.fr/users/roussel/herschel/} \citep{rou12}.

For photometry, the Herschel data were processed with the high-pass filtering method described in \citet{fis10}, using version 8.0, build 248, of HIPE, the Herschel Interactive Processing Environment \citep{ott10}.  V2775 Ori is unresolved at both PACS wavelengths.  We obtained aperture photometry using a 9.6\arcsec\ radius at 70~\micron\ and a 12.8\arcsec\ radius at 160~\micron\ with subtraction of the median signal in a background annulus extending from the aperture limit to twice that value in both channels.  The results were divided by measurements of the encircled energy fractions in these apertures provided by the {\tt photApertureCorrectionPointSource} task in HIPE, adjusted for the fact that our close-in sky subtraction removes 3--4\% of the flux in each point-spread function.  These aperture corrections are 0.733 at 70 \micron\ and 0.660 at 160 \micron.  The error estimate is dominated by calibration uncertainties, which we estimate to be 5\% at both wavelengths.

\subsection{APEX Imaging and Photometry}

In 2011 September, V2775 Ori was observed with the SABOCA and LABOCA instruments on APEX, the Atacama Pathfinder Experiment.   SABOCA, the Submillimetre APEX Bolometer Camera \citep{sir10}, operates at 350 \micron.  It was used to observe two fields containing the source, one on 2011 September 16 and one on 2011 September 18.  LABOCA, the Large APEX Bolometer Camera \citep{sir09}, operates at 870~\micron.  It also was used to observe two fields containing the source, one on 2011 September 22 and one on 2011 September 26.  The two fields at each wavelength were combined.  Cutouts from the APEX fields appear in the bottom row of Figure~\ref{f.farir}.

For LABOCA, the data were first put on an approximate flux density scale using a standard factor to convert counts to flux density. Atmospheric absorption was accounted for with atmospheric zenith opacities derived from skydips taken every one to two hours. Finally, planets and secondary flux density calibrators (observed at least once per hour) were used to establish corrections to the standard flux density conversion and to account for imperfection in the opacity determination from the skydips. This yields a flux density calibration that is accurate to about 10\%.  We quote a larger 40\% uncertainty for V2775 Ori, as it is unclear how much the surrounding cloud/filament and the two other sources in the field contribute to the 870 \micron\ flux density.

For SABOCA, the procedure was similar, except that we used the atmospheric model directly (with precipitable water vapor as measured by the APEX radiometer at the time of the actual observation as input) to get the zenith opacity. This has the advantage that the opacity is derived for the time and elevation at which the observation was actually taken, not from a nonsimultaneous skydip.  Nevertheless, the scatter of calibration factors remains substantial, as even relatively small changes in precipitable water vapor cause a fairly large change in atmospheric attenuation at 350 \micron, where the opacity remains high even in very good conditions. From the scatter of the calibration factors, the flux densities cannot be derived to better than 30--40\% for SABOCA outside of the very best conditions, which we did not have for V2775 Ori.  We thus quote a 40\% uncertainty for SABOCA as well.

At both wavelengths, we report the flux densities without sky subtraction in circular apertures of diameter equal to the FWHM of the instrument beam: 7.3\arcsec\ at 350~\micron\ and 19\arcsec\ at 870~\micron.

\subsection{IRTF/SpeX Spectroscopy\label{s.spex}}

We observed V2775 Ori with SpeX \citep{ray03} at the Infrared Telescope Facility (IRTF) on 2011 January 27, 29, 31, and February 1.  We used the short-wavelength cross-dispersed (SXD) mode, selecting a 0.3$\arcsec\times15\arcsec$ slit and obtaining spectra that extend from 1.0 to 2.4~$\micron$ at a resolving power $R=2000$.  (Due to the extreme redness of the object, no continuum could be extracted over the 0.8 to 1.0~$\micron$ range normally accessible in SXD mode.)  Total exposure times were 36 minutes on 29 January and 32 minutes on the other nights, yielding a continuum $S/N\sim20$ at the one-micron $Y$ band, increasing to about 120 at $K$.

To facilitate the subtraction of sky emission lines, the observing sequence for a given object consisted of multiple 120 s exposures.  After the first and third exposures of a four-exposure sequence, the telescope was nodded by 7.5$\arcsec$ (an ABBA pattern) such that the object remained in the slit.  The normal A0 stars HD 35505 and HD 45137, near the target and at similar airmass, were observed to allow the removal of atmospheric absorption lines and provide an approximate flux calibration.

The data were reduced with Spextool \citep{cus04}, an IDL package containing routines for dark subtraction and flat fielding, spectral extraction, wavelength calibration with arc lamp lines, and the combining of multiple observations from a nodding sequence.  The correction for atmospheric absorption lines was handled by {\em xtellcor} \citep{vac03}, an IDL routine that divides the target spectrum by a calibrator.  If the calibrator is an A0 star, then its spectrum consists to a good approximation only of hydrogen lines and atmospheric absorption lines.  Division by such a calibrator, if observed at approximately the same airmass as the target, removes the target's atmospheric absorption lines.  The hydrogen lines in the calibrator were removed by fitting a scaled model of Vega; inspection of the spectra of V2775 Ori before and after the telluric correction confirm that this process does not generate spurious hydrogen emission or absorption.  After telluric correction, the merging of the six orders covered in the SXD mode and the removal of remaining bad pixels were accomplished with separate programs in the Spextool package.

The uncertainty in the SpeX flux calibration is dominated by different slit losses for the target and the calibrator.  Over the four nights, we measured some variation in the $J$, $H$, and $K$ magnitudes: $12.37~{\rm mag}<J<12.68~{\rm mag}$, $9.70~{\rm mag}<H<10.03~{\rm mag}$, and $8.11~{\rm mag}<K<8.42~{\rm mag}$.  The $J-H$, $H-K$, and $J-K$ colors, however, were remarkably consistent from night to night, varying by at most 0.04 magnitudes.  The lack of color change points to wavelength-independent slit losses, not intrinsic variability, and in the SED of V2775 Ori we assume that the absolute flux of the source during the SpeX run is well represented by the mean of the four spectra.
 
\subsection{Spitzer/IRS Spectroscopy}

We observed V2775 Ori (target 8570200-8276310) on 2007 March 12 with the Spitzer Infrared Spectrograph (IRS; \citealt{hou04}). The observations were made with the two low-spectral-resolution IRS modules (SL from 5.2 to 14.5 \micron\ and LL from 14.0 to 38.0 \micron; $\lambda/\Delta\lambda=60-120$) in staring mode.  The spectrum was generated from the Spitzer Science Center (SSC) S18.7 pipeline basic calibrated data with the IRS instrument team's SMART software package \citep{hig04}.  To prepare the data for extraction, we first replaced the permanently bad and ``rogue'' pixels' flux values with those interpolated from neighboring functional pixels.  Background emission was removed by subtracting the opposite nod positions.  The reduced two-dimensional data were extracted using a column with a width that was matched to the IRS point-spread function.

We then used the same method to extract spectra of the calibrators Markarian 231, $\alpha$ Lac, and $\xi$ Dra.  We divided template spectra of the calibrators by their extracted spectra at the two nod positions to create relative spectral response functions (RSRFs).  The extracted orders of V2775 Ori at both nod positions were multiplied by the RSRFs, and the resulting nod-position spectra were averaged to obtain the final spectrum. The spectral uncertainties are estimated to be half the difference between the independent spectra from each nod position; the median signal-to-noise ratio of the spectrum is 150.

\subsection{Palomar/Triplespec Spectroscopy of FU Ori\label{s.tspec}}

For comparison with the SpeX spectrum of V2775 Ori in Section~\ref{s.nirspec}, we obtained a spectrum of FU Ori with Triplespec \citep{her08} at the Hale 200-inch telescope of Palomar Observatory on 2012 January 9.  We obtained four consecutive one-minute exposures of FU Ori for a continuum $S/N\sim400$ in the $Y$ band, increasing to about 800 at $K$.  The data were reduced with a version of Spextool modified for Triplespec by M. Cushing, following the same steps as for the SpeX observations of V2775 Ori and using HD 34317 as a telluric calibrator.  The spectrum extends from 1.0 to 2.5~$\micron$ at a resolving power $R=2700$.

\section{IMAGE OVERVIEW}

Here we present a brief overview of the new images described above.   In the HST image, V2775 Ori and an associated reflection nebula are visible, as well as [CTF93] 216 (2MASS J05424706-0817069; HOPS 221) and its associated nebulosity, 39\arcsec\ to the southwest.  The latter is an embedded protostar with a redder near-to-mid-IR SED than V2775 Ori, indicating a denser, less-evolved envelope \citep{car12}.

The near-IR image also shows a point source 11\arcsec\ to the northwest of V2775 Ori, 2MASS J05424814-0816250.  It does not have an IR excess \citep{meg12}.  The 2MASS colors and magnitudes are most consistent with a foreground dwarf of class M0 ($d=140~{\rm pc}$, extinction $A_K=0.30~{\rm mag}$) or a foreground dwarf of class K2 ($d=205~{\rm pc}$, $A_K=0.45~{\rm mag}$).  If the star were a background giant, it would be a G8 star at $2630~{\rm pc}$ with a extinction $A_K=0.30~{\rm mag}$, prohibitively low if it is viewed through a molecular cloud complex.  The source may also be a pre--main-sequence companion to V2775 Ori at a separation $\ge 4600~{\rm AU}$.

As in Figure~\ref{f.wfc3}, V2775 Ori appears at the center of the Herschel and APEX images shown in Figure~\ref{f.farir}, and [CTF93] 216 appears in the southwest.  A source newly identified by the HOPS team, invisible in HST, 2MASS, Spitzer/IRAC, and 24~\micron\ Spitzer/MIPS images, appears 25\arcsec\ to the northeast of V2775 Ori at 70~\micron\ and longer wavelengths.  Due to its extremely red SED, this is likely a cold source, more embedded and less evolved than the protostars detected in Spitzer 3--24 \micron\ images.  It is part of a sample of such objects discovered in the HOPS images (Stutz et al., in prep.).  During all HOPS observations the LABOCA beam was very close to circular; the elongation in the 870 \micron\ image is not due to a noncircular beam, but rather to blending with the source to the northeast and perhaps dust between the two sources.

\section{TIMING OF THE OUTBURST}

The Spitzer/IRS spectrum of 2007 March 12 shows flux increases from the 2004 and 2005 Spitzer photometry at 8.0 and 24~\micron\ by factors of 15.7 and 13.4, respectively.  We thus conclude that the outburst took place between 2005 April and 2007 March.  The source faded between the acquisition of the IRS spectrum in 2007 and the second imaging by MIPS in 2008 November, with a 45\% reduction in the 24 \micron\ flux over this period.

Our analysis places the outburst earlier than the range of dates proposed by \citet{car11}.  In their Figure~3, their archival IRS spectrum from the same date is plotted with a median flux of $\sim 10^{-10}$ erg cm$^{-2}$ s$^{-1}$, making it consistent with the 2004 and 2005 photometry.  They accordingly place the outburst {\em after} the acquisition of the IRS spectrum, i.e., between 2007 March and the 2008 November observation by MIPS.  Our version of the same IRS spectrum has a median flux of $\sim 10^{-9}$ erg cm$^{-2}$ s$^{-1}$, in agreement with the spectra obtainable via the Cornell Atlas of Spitzer/IRS Sources\footnote{http://cassis.astro.cornell.edu/atlas/} (CASSIS; \citealt{leb11}) or the Spitzer Heritage Archive,\footnote{http://sha.ipac.caltech.edu/} which were reduced independently with different software and different calibrators and have calibration uncertainties of order 1\%.

\section{SED ANALYSIS}

To analyze the SED evolution of V2775 Ori, we separate the data into four epochs.  Epoch 1 is the 2MASS epoch (1998), Epoch 2 is the first Spitzer epoch (2004--2005), and Epoch 3 is the IRS epoch (2007 March, after the outburst).  Since the 24 \micron\ MIPS photometry from 2008 November and the 22 \micron\ WISE photometry from 2010 March are nearly identical, the SED appears to have been relatively stable since late 2008.  We thus define all data obtained from 2008 November to 2011 September as Epoch 4.  The Epoch 4 SED is well sampled from 1.0 to 870 \micron, making it suitable for modeling with a radiative transfer code.   Even without rigorous modeling, the wide, flat shape of the Epoch 4 SED is informative.  The ratio of $\lambda F_\lambda$ at $K$ to that at 70 \micron\ is near unity (0.68).  Very dense envelopes absorb almost all near-IR flux, and the most tenuous envelopes emit very little far-IR flux, pointing to a moderate envelope density for the object.

\begin{figure*}
\plotone{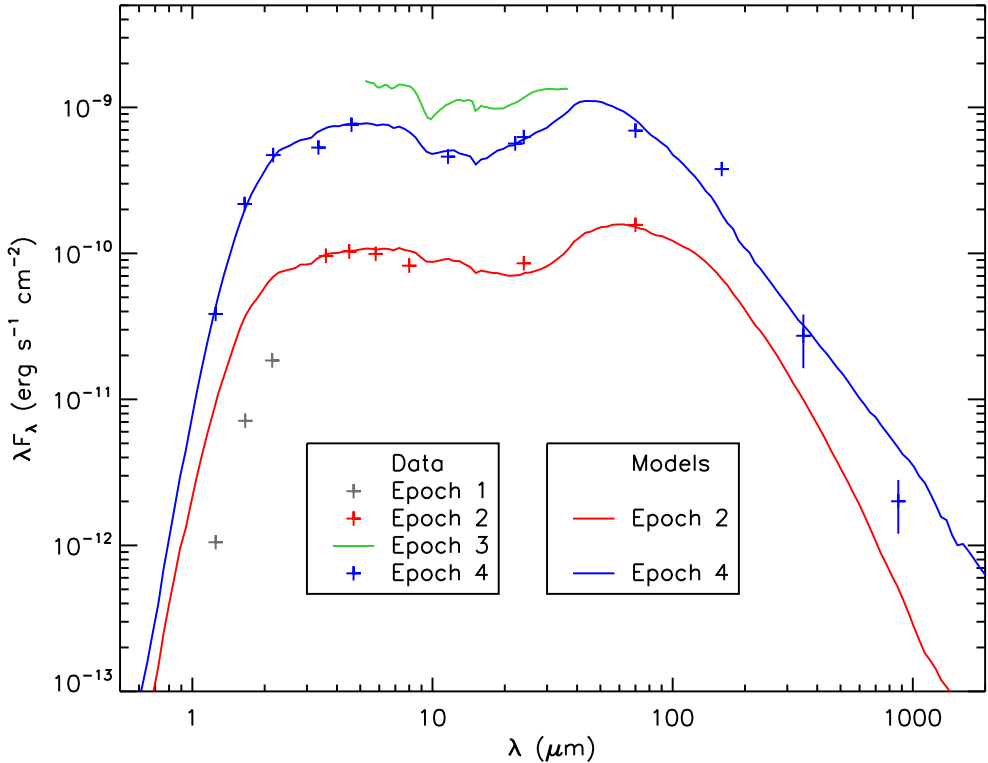}
\figcaption{Comparison of the evolving SED of V2775 Ori to results from a grid of 3,600 model SEDs calculated with the radiative transfer code of \citet{whi03}.  The post-outburst (Epoch 4) model is the best fit to the relevant data, while the pre-outburst (Epoch 2) model has the same envelope and disk properties as the post-outburst model but a lower luminosity.  The modeled luminosity increased from $4.5~L_\sun$ in Epoch 2 to $28.1~L_\sun$ in Epoch 4.  The source of each data product can be determined from Figure~\ref{f.sed}.\label{f.fit}}
\end{figure*}

\subsection{Fitting with a Grid of 3,600 Models}

To more precisely estimate the source's properties, we fit the Epoch 4 SED with a grid of 3,600 models created with the Monte Carlo radiative transfer code of \citet{whi03}.  The code contains a central luminosity source, a luminous accretion disk with power-law radial density profile and scale height, an envelope with a density that follows the rotating collapse solution of \citet{ter84}, and a bipolar envelope cavity.  For protostars, the SEDs are most sensitive to the total luminosity and the envelope density.  The envelope density law can be scaled to match envelopes of varying average densities; we quantify this scaling with a reference envelope density $\rho_1$, the density at 1~AU in the limit of no rotation.  For a \citeauthor{ter84}\ envelope, \begin{equation}\rho_1=7.4\times10^{-15}\!\left(\frac{\dot{M}_{\rm env}}{10^{-6}~M_\sun~{\rm yr}^{-1}}\right)\!\!\!\left(\frac{M_*}{0.5~M_\sun}\right)^{-1/2}\!\!\!\rm{g~cm}^{-3}\end{equation} \citep{ken93}, where $\dot{M}_{\rm env}$ is the infall rate of the envelope onto the disk and $M_*$ is the mass of the central source.  Thus, a reference envelope density can be converted to an envelope infall rate with an assumed central mass.

The grid is a refinement of the one described in \citet{ali10}.  The central source has fixed properties, with a radius of $R_*=2.09~R_\sun$, a mass of $M_*=0.5~M_\sun$, and an effective temperature of $4000~{\rm K}$.  In embedded low-mass protostars, reprocessing of luminosity by the envelope generally destroys information about the central star, so this simplification does not reduce the range of SED shapes in the grid.  The envelope dust properties are the same as those described by \citet{tob08} except that carbon is amorphous instead of in the form of graphite; the grains are larger than a standard interstellar dust model and are thus more suitable for star-forming regions.  The protostellar properties vary across the grid as follows:  There are three disk radii of 5, 100, and 500 AU.  There are three envelope radii of 5,000, 10,000, and 15,000 AU.  There are four disk accretion rates that control the total luminosity, ranging from $10^{-8}~M_\sun$~yr$^{-1}$ (1.06 $L_\sun$) to $10^{-5}~M_\sun$~yr$^{-1}$ (56.3 $L_\sun$) in steps of one dex.  There are five cavity opening half-angles extending from 5$^\circ$ to 45$^\circ$ in steps of 10$^\circ$.  Finally, there are twenty envelope infall rates taking on values of 0 (disk-only) and a range from $5\times10^{-8}~M_\sun$~yr$^{-1}$ to $1\times10^{-3}~M_\sun$~yr$^{-1}$ in approximately equal logarithmic steps.  (With a central star of mass $0.5~M_\sun$, the reference envelope densities $\rho_1$ for the models with envelopes extend from $3.7\times10^{-16}~{\rm g~cm}^{-3}$ to $7.4\times10^{-12}~{\rm g~cm}^{-3}$.)  Each model can be viewed from one of ten angles, resulting in a grid of 36,000 SEDs.

During fitting, a model SED can be modified in two ways to improve the fit.  First, it can be multiplied by a constant between 0.3 and 3.3 (corresponding to $\pm1.3$~mag), which allows the source luminosity to fall along a continuum instead of the four discrete values represented in the grid.  The wavelength of the peak flux from an optically thick envelope scales as the luminosity to only the $-1/12$ power \citep{ken93}, so for factors sufficiently close to unity, a change in luminosity does not significantly change the shape of the SED.  Second, the model SED can be subjected to foreground reddening applied with the law of \citet{mcc09}, which was determined from observations of nearby star-forming regions.

For the fitting of V2775 Ori, we used $J$, $H$, and $K_s$ photometry derived from the SpeX spectrum, the WISE photometry, the 2008 MIPS photometry, and the PACS, SABOCA, and LABOCA photometry.  To help constrain the fit, we also used flux measurements at 8, 9.7, 12, 18, and 32 \micron\ that we derived from the IRS spectrum after scaling it down to match the MIPS point at 24 \micron.  This assumes no evolution in the IRS spectral shape between Epochs 3 and 4; since the resulting fits are consistent with the Epoch 4 photometry, this scaling appears to be reasonable.

The fitting for the APEX data is aperture dependent.  At most wavelengths, V2775 Ori is pointlike, and we compare observed fluxes measured in a finite aperture and corrected to infinity to the total model flux.   In the APEX maps, V2775 Ori is slightly resolved.  At these wavelengths, we compare observed fluxes in an aperture with diameter equal to the FWHM of the instrument beam to model fluxes in the same aperture.

Since the deviations among the SEDs in the model grid are typically larger than the measurement uncertainties and the same is expected for the data relative to the models, we eschew the usual $\chi^2$ analysis of fit quality.  Instead, we measure the logarithmic deviation of the observed SED from the models using a statistic \begin{equation}R=\sum_{i=1}^N\left[w_i\left| \ln \left(F_{\lambda_i{\rm, o}}/F_{\lambda_i{\rm, m}}\right)\right|\right]/N.\end{equation}  Here $N$ is the number of data points, $F_{\lambda_i{\rm, o}}$ is the observed flux at each wavelength $\lambda_i$, $F_{\lambda_i{\rm, m}}$ is the model flux at each wavelength $\lambda_i$, and $w_i$ is the inverse of the approximate fractional uncertainty in each data point.  For small deviations, $R$ is equal to the distance of the model from the data in units of the fractional uncertainty.  Minimizing $R$ parallels the common process of comparing the model and observed SEDs by eye on a plot of $\log (\lambda F_\lambda)$ versus $\log \lambda$.

Using the measured uncertainties in Table~\ref{t.photometry2} as a guide, we assigned the WISE, Spitzer, and 70~\micron\ Herschel points $w_i=1/0.05$, the SpeX points $w_i=1/0.10$, and the APEX points $w_i=1/0.40$.  The weight for the 160~\micron\ Herschel point, $w_i=1/0.10$, is half that of the 70~\micron\ point.  While their formal errors are similar, the 160~\micron\ point can be influenced by external dust that may not be well modeled by an isolated spherical envelope.

The best-fit model for the Epoch 4 SED is shown in Figure~\ref{f.fit}.  It has $R=1.84$ per data point and $\chi^2=5.53$ per data point.  (This is also the best model by $\chi^2$ ranking.)  The model has a reference envelope density of $7.4\times10^{-15}~\rm{g~cm}^{-3}$.  With this reference density, the envelope mass inside the outer radius of $10^4$~AU is $0.09~M_\sun$.  The cavity opening half-angle is $25^\circ$, the inclination angle is $49^\circ$ ($0^\circ$ is face-on), and the foreground extinction is 1.14 mag at $K$.  The bolometric luminosity found by integrating under the Epoch 4 photometry is $16.4~L_\sun$, but correcting for the modeled foreground extinction, which is responsible for the majority of the correction, and inclination \citep{whi03} gives an intrinsic luminosity of $28.1~L_\sun$.

While the fit is satisfying in the near-to-mid-IR, the model underpredicts the 160~\micron\ flux and overpredicts the 870~\micron\ flux; i.e., the observed SED is steeper in the far IR than the model one.  A likely factor in the discrepancy is the large beam sizes for the 160~\micron\ and 870~\micron\ observations.  These points may be influenced by the dust distribution outside the radius modeled as an isolated collapsing envelope.  A different dust opacity law could also improve the fit, and variability between the 2008 Herschel observations and the 2011 APEX observations could be a factor.  Performing the fit again and excluding these two points results in the same best model, with $R=1.38$ instead of 1.84, so the mismatches at these two points are unlikely to have a strong influence on the inferred parameters.

To estimate the ranges of model parameters that may apply to V2775 Ori post-outburst, we measure $R$ for the full grid of models against the best-fit model; this is denoted $R_{\rm model}$ and is 0 for the best-fit model.  We consider the models that have $R_{\rm model}=1.38\pm1$; i.e., they are as far from the best-fit model as the data (when removing the discrepant 160 and 870 \micron\ points from consideration).  The following ranges are the standard deviations of the model parameters in this group.  The luminosities range from 22.0 to $34.2~L_\sun$, the reference envelope densities range from 2.3 to $23\times10^{-15}~\rm{g~cm}^{-3}$, the cavity opening half-angles range from 19 to 31$^\circ$, the inclination angles range from 34 to 65$^\circ$, and the foreground extinctions range from 0.89 to 1.39 mag at $K$.

The ranges of luminosities and inclination angles reflect the usual degeneracy in these quantities, where an inclination angle closer to edge-on implies the system luminosity is larger than observed due to extinction by the disk or denser equatorial regions of the envelope (see Fig.~10 of \citealt{whi03}).  The mean intrinsic luminosity of the allowed models at each inclination angle increases monotonically from the most face-on permitted angle to the most edge-on permitted angle.  Our high-resolution near-IR image (Fig.~\ref{f.wfc3}) lacks the dark central lane indicative of an edge-on orientation, but it does not distinguish well among the intermediate inclination angles indicated by SED modeling.   While the best-fit SED indicates an inclination and luminosity in the middle of the quoted ranges, the true luminosity may differ by 30 to 40\% in either direction.

Independent fits to the pre-outburst SED are not as well constrained as fits to the Epoch 4 SED due to the smaller number of photometric data points.  (We find no entry for this source in the Submillimetre Common User Bolometer Array [SCUBA] legacy catalog of \citealt{dif08}.)  We instead adopt the best-fit Epoch 4 model and decrease the luminosity to minimize $R$ for the Epoch~2 SED.  This requires a step down in the disk accretion rate parameter of the model, so it is not simply a multiplication of the Epoch 4 model SED by a constant.  The best fit to the Epoch 2 data according to this prescription, shown in Figure~\ref{f.fit}, has an intrinsic luminosity of $4.5~L_\sun$.  Its $R$ of 1.65 is similar to the $R$ of the best Epoch 4 model; the fact that the fit is of the same quality suggests that our assumption of no evolution in the envelope properties is reasonable.  This fit overpredicts the Epoch 1 (2MASS) fluxes, which were obtained 5.5 years before the earliest Epoch 2 data, which may indicate a slow rise in luminosity before the major outburst similar to that of V2493 Cyg \citep{mil11}.  We thus find an increase in the total luminosity of the source from 4.5 to $28.1~L_\sun$, a factor of 6.2, between Epoch 2 and Epoch 4.  (The peak luminosity was reached between these epochs, as demonstrated by the Epoch 3 IRS spectrum.)

Recently, a source with an SED similar to that of V2775 Ori but somewhat wider and flatter between 2 and 160~\micron\ was modeled by \citet{ada12} as a binary in which one object is a source with only a disk and the other is a protostar still deeply embedded in an envelope.  From our high-resolution near-IR image, we see no evidence for binarity, but we cannot exclude the possibility.  This scenario is worth exploring in future work on V2775 Ori.

\subsection{Accretion Rates from Stellar Properties}

The rate at which the disk accretes onto the star and the rate at which the envelope falls onto the disk, which are not necessarily equal, cannot be specified precisely without knowledge of the central source properties.  The disk accretion rate is $\dot{M}_{\rm disk}=\epsilon L_{\rm acc}R_*/GM_*$, where $G$ is the gravitational constant, and $L_{\rm acc}$ is the luminosity due to accretion (the total luminosity minus the luminosity of the central source).  The factor $\epsilon$ is close to unity and depends on the details of the accretion process.  If the accretion luminosity is generated as gas free-falls from the inner radius of the (gas) accretion disk $R_T$ to the stellar surface, then $\epsilon=1/\left(1-R_*/R_T\right)$.  For $R_T=5~R_*$, representative of the range found by \citet{mey97} for accreting T Tauri stars, $\epsilon=1.25$.  The following findings would not change substantially for other $\epsilon$ close to unity.

\citet{car11} modeled their 2010 near-infrared spectrum to find an effective temperature $T_{\rm eff}=3200~{\rm K}$ for the central source.  The luminosity of the central source is ambiguous, as it depends on the fraction of the total luminosity assumed to be due to accretion.  Interpolating the pre--main-sequence models of \citet{sie00}, a star of this effective temperature has a mass of about $0.24~M_\sun$ and a radius that decreases from $4.4~R_\sun$ to $1.9~R_\sun$ as its age increases from $10^4$ to $10^6$ years.  Given the temperature and range of radii, the luminosity of the star decreases from $1.8~L_\sun$ to $0.3~L_\sun$ over this time.

From our SED modeling, the total pre-outburst (Epoch 2) luminosity is $4.5~L_\sun$, meaning the pre-outburst accretion luminosity ranges from 2.7 to $4.2~L_\sun$ after subtracting the range of central luminosities given above.  The total post-outburst (Epoch 4) luminosity is $28.1~L_\sun$, meaning the post-outburst accretion luminosity ranges from 26.3 to $27.8~L_\sun$.

For the assumed stellar mass, gas truncation radius, range of stellar radii, and ranges of accretion luminosities, the pre-outburst disk accretion rate is between 1.3 and $2.0\times10^{-6}~M_\sun~\rm{yr}^{-1}$, and the post-outburst disk accretion rate is between $8.8\times10^{-6}$ and $1.9\times10^{-5}~M_\sun~\rm{yr}^{-1}$, with the largest accretion rates applying to a system of age $10^4$~yr and the smallest applying to a system of age $10^6$~yr.  Depending on age, the disk accretion rate increased by a factor of 7 to 10 between Epoch 2 and Epoch 4.  (The peak accretion rate occurred between these epochs, as demonstrated by the Epoch 3 IRS spectrum.)  The range of post-outburst disk accretion rates is about an order of magnitude lower than the canonical disk accretion rate of $10^{-4}~M_\sun~\rm{yr}^{-1}$ for FU Orionis objects \citep{har96}.

Given the modeled reference envelope density of $\rho_1=7.4\times10^{-15}~\rm{g~cm}^{-3}$, the envelope infall rate depends only on the central mass (Equation~1).  The envelope infall rate is $6.9\times10^{-7}~M_\sun~\rm{yr}^{-1}$ for a central mass of $0.24~M_\sun$.

For the range of stellar parameters explored, the envelope infall rate is less than the pre- and post-outburst disk accretion rates.  Further, the envelope mass inside the $10^4$~AU outer radius ($0.09~M_\sun$) is less than the estimated mass of the central star ($0.24~M_\sun$), indicating that most of the envelope has been accreted.  While the appearance of a 10 \micron\ silicate absorption feature in its IRS spectrum indicates membership in the embedded Category 1 of \citet{qua07}, the low envelope mass and infall rate imply that V2775 Ori is approaching the end of its envelope-embedded phase.

\section{NATURE OF THE OUTBURST}

Among young stars with high accretion rates, FU Orionis sources are only the most extreme variety, with high luminosities and spectral features that distinguish them from normal active accretors.  An increase in the luminosity or accretion rate of a source does not necessarily place it in the FU Orionis regime.  Our post-outburst near-IR image of V2775 Ori  (Fig.~\ref{f.wfc3}) shows a broad reflection nebula, typical of FU Orionis objects \citep{har96}.  To further characterize its outburst, we examine the evolution of its luminosity and the appearance of its near-infrared spectrum.

\subsection{Luminosity Evolution}

Between early 2005 (Epoch 2) and early 2007 (Epoch 3), V2775 Ori experienced an increase in its 24 \micron\ flux density of a factor of 13.4, or 2.8 magnitudes.  The initial rise time was thus at most two years.  By late 2008 (Epoch 4), the 24 \micron\ flux had diminished by 45\%.  We can make robust estimates of the total luminosity in Epochs 2 and 4 due to the relatively dense wavelength sampling of the SEDs at those times (Section 5):  The source increased in luminosity from $4.5~L_\sun$ in Epoch 2 to $28.1~L_\sun$ in Epoch 4.  The change in total luminosity between these epochs was roughly proportional to the change in 24~\micron\ flux density: the luminosity was 6.2 times larger and the 24~\micron\ flux density was 7.3 times larger.  If this proportionality also held from Epoch 3 to Epoch 4, then the IRS spectrum reflects an Epoch 3 luminosity of $\sim51~L_\sun$.  Both the change in luminosity and the peak luminosity are within but toward the low end of the range for FU Orionis objects \citep{rei10}.

After the 45\% drop in the 24 \micron\ flux density between 2007 March and 2008 November, the object showed evidence of subsequent slight dimming.  In the $K$ band, where the photometric time-coverage is best, \citet{car11} reported a $K_s$ magnitude of 7.98 in 2009 February, a fading of 0.14 mag by the middle of 2010 October, and a fading of another 0.13 mag by the end of that month.  Our $K$-band spectra of 2011 January have an average $K_s=8.23~{\rm mag}$, nearly equal to the $K_s=8.25~{\rm mag}$ reported by \citet{car11} for the end of 2010 October.  It is unclear whether this dimming of 0.25 mag at $K_s$ over two years is the beginning of a slow decline to the pre-outburst state or a temporary change; more observations are required.

\subsection {Near-Infrared Spectra\label{s.nirspec}}

Young stellar objects with high accretion rates can have the near-IR spectra of either normal active accretors or of FU Orionis objects. \citet{con10} presented near-IR spectra of wavelength range and resolution similar to those analyzed here of 110 embedded young stars, 10 of which had FU Orionis-like spectra, and they characterized the differences between the spectra of normal active accretors and FU Orionis objects.  Normal active accretors have narrow metal absorption lines typical of cool young stars but sharply reduced in strength by excess continuum emission, CO lines in the $K$ band that are either in emission or weakly in absorption, and strong emission lines of atomic hydrogen.  FU Orionis objects, in contrast, have broad metal absorption lines not easily distinguished at moderate spectral resolution, H$_2$O and CO absorption features that are deeper than expected from a cool stellar photosphere, and no atomic hydrogen emission. 

\begin{figure*}
\resizebox{\hsize}{!}{\includegraphics{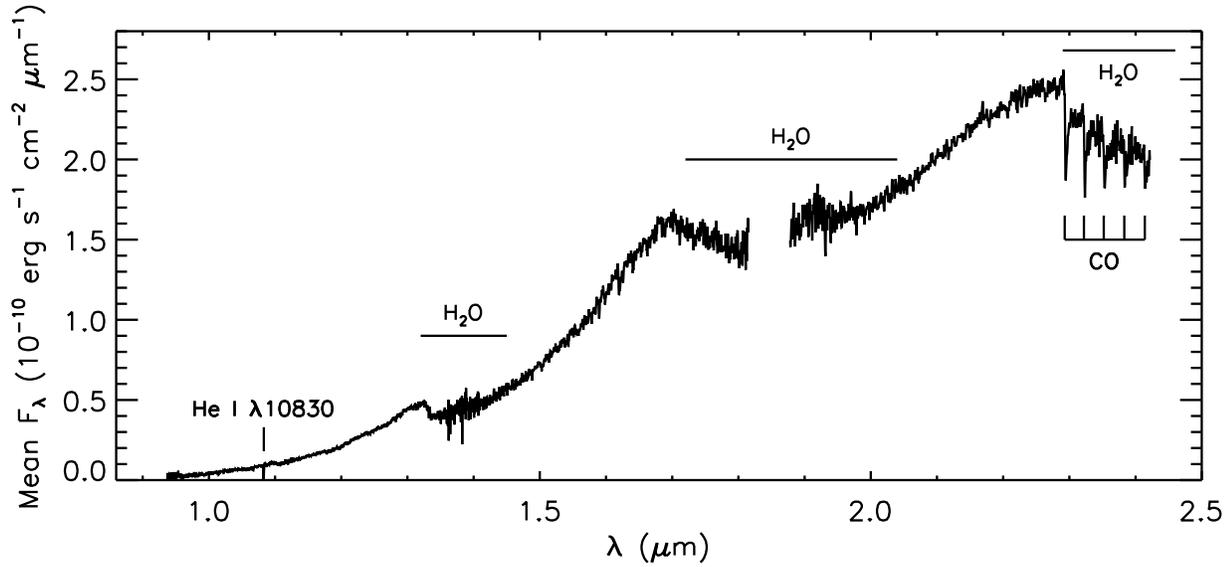}}
\figcaption{IRTF/SpeX spectrum of V2775 Ori.  This is the average of four spectra obtained in 2011, on January 27, January 29, January 31, and February 1.  Important spectral features are marked.  Much of the noise in the water bands is due to imperfect removal of strong telluric water features.\label{f.spex}}
\end{figure*}

In the near-IR spectra acquired by us and \citet{car11} between 2009 February and 2011 February, V2775 Ori displayed characteristics of an FU Orionis object.  \citet{car11} presented two spectra: an NTT/SofI spectrum extending from 1.5 to 2.4~\micron\ at $R=1000$ obtained on 2009 February 13 and a TNG/NICS spectrum extending from 1.1 to 2.4~\micron\ at $R=600$ obtained on 2010 October 13.  Just 15 weeks later, we obtained IRTF/SpeX spectra with similar wavelength coverage and resolution (1.0 to 2.4 \micron\ at $R=2000$) on four nights of a six-night run, between 2011 January 27 and February 1.  We see no evidence for variability over our four SpeX spectra, so we analyze the mean spectrum, shown and annotated in Figure~\ref{f.spex}.  

\begin{figure*}
\resizebox{\hsize}{!}{\includegraphics{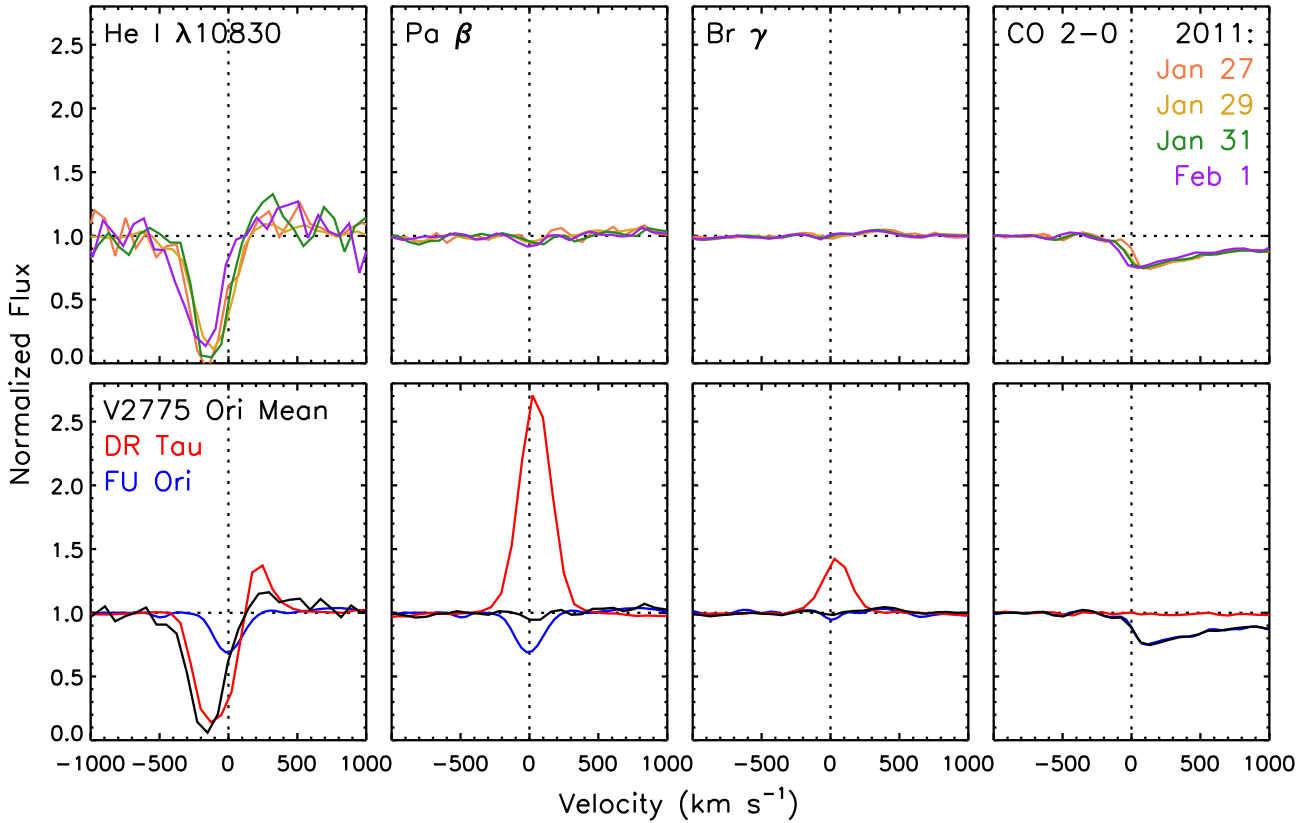}}
\figcaption{Continuum-normalized profiles of \helium, the \ion{H}{1} lines Pa$\beta$ and Br$\gamma$, and the CO 2-0 bandhead at 2.29 \micron. {\em Top row:} IRTF/SpeX profiles of V2775 Ori from each of the four nights it was observed.  There is no evidence for variability from night to night. {\em Bottom row:} The mean IRTF/SpeX profiles of V2775 Ori are compared to those of DR Tau from the same instrument (2006 Nov 27) and those of FU Ori from Palomar/Triplespec (9 Jan 2012).\label{f.lines}}
\end{figure*}

Figure~\ref{f.lines} shows four lines that form in the circumstellar environment.  The top row shows \helium, the \ion{H}{1} lines Pa$\gamma$ and Br$\gamma$, and the 2.29~\micron\ CO 2-0 line from the four SpeX spectra of V2775 Ori.  As is the case for the continuum shape, we see no compelling evidence for line variability in the four spectra.  The bottom row compares mean profiles from V2775 Ori to profiles of two other objects.  First, DR Tau is a T Tauri star with a large disk accretion rate ($\dot{M}_{\rm disk}=5\times10^{-7}~M_\sun$~yr$^{-1}$; \citealt{muz98}).  Its spectrum was obtained with SpeX on 2006 November 27 and is discussed in \citet{fis11}.  We use it to represent the case of normal active accretors.  Second, FU Ori is the prototype of the FU Orionis class.  It is interpreted to have such a high disk accretion rate that the usual spectroscopic signatures of normal active accretors are absent, marking the transition to another regime of accretion physics.  Its spectrum was obtained with Triplespec at Palomar Observatory on 2012 January 9 (Section~\ref{s.tspec}).

All spectra of V2775 Ori obtained between 2009 and 2011 show the deep CO absorption and broad H$_2$O absorption typical of FU Orionis objects.  The equivalent width of the 2.29~\micron\ CO 2-0 line in our SpeX spectrum is 10.0~\AA.  In Figure 8 of \citet{con10}, this equivalent width places V2775 Ori on the dividing line between Region B (weaker absorption in normal stellar photospheres) and Region C (stronger absorption in FU Orionis objects).  The CO line in FU Ori has the same equivalent width; it is virtually indistinguishable from the V2775 Ori profile in Figure~\ref{f.lines}.  This equivalent width is not typical of normal actively accreting stars; DR Tau has no detectable CO 2-0 feature, and the CO lines of normal active accretors are often in emission.

In normal accreting young stars, the \helium\ line traces infall and outflow within about 0.1 AU of the stellar surface \citep{edw06,kwa07,fis08}.  The second paper of that series showed that deep, broad, blueshifted \helium\ absorption can be explained by scattering in a wind that accelerates radially from the source of the one-micron continuum, such that its entire acceleration region is seen in projection against the continuum source.  In the objects studied in that work, the one-micron continuum mostly comes from the star, and the wind probed by the deep, broad, blueshifted helium absorption is a stellar wind.  In FU Orionis objects, the one-micron continuum is mainly due to the disk, and such profiles would signal a disk wind.

Here we present the first detections of \helium\ in V2775 Ori.  It has a deep, blueshifted absorption profile, extending from roughly zero velocity to beyond $-400$~km~s$^{-1}$, with equivalent width 10.6~\AA.  There is weak emission such that the net equivalent width is 8.0~\AA.  In Figure~\ref{f.lines}, DR Tau has a similar absorption profile with equivalent width 9.7~\AA\ (net equivalent width 7.3~\AA), and FU Ori shows a narrower, shallower, profile with no convincing blueshift, no emission, and total equivalent width 2.8~\AA.  Since both FU Orionis objects and normal active accretors can show a large range of blueshifted \helium\ morphologies, as seen here and in \citet{con10}, this transition alone is not a discriminant of FU Orionis-like activity.

The near-IR hydrogen lines of T Tauri stars are often used to estimate accretion luminosities in young stellar objects \citep{muz98}.  Indeed, DR Tau has strong Paschen and Brackett lines as expected from its accretion luminosity of 2.7 $L_\sun$ independently determined from its optical/UV continuum emission \citep{cal98}.  The equivalent widths of its Pa$\beta$ and Br$\gamma$ lines seen in Figure~\ref{f.lines} are 18.4~\AA\ and 8.6~\AA, respectively.  In FU Ori and other stars of its class, the hydrogen lines are weak or absent even though the accretion luminosity is very high (hundreds of solar luminosities for FU Ori; \citealt{har96}).  We see no atomic hydrogen emission stronger than 0.8~\AA\ in our spectra of V2775 Ori.\footnote{\citet{car11} reported Br$\gamma$ emission of $\sim3.6$ \AA\ in their post-outburst spectra.  When we obtained their 2009 raw spectra from the ESO archive and reduced them, we saw no emission feature.}  With $K_s=8.23~{\rm mag}$ in our averaged spectrum, $A_K=1.14$ from our SED modeling, and a distance of 420 pc, the Br$\gamma$ luminosity is thus less than $2.7\times10^{-4}~L_\sun$.  If the \citet{muz98} relationship were to hold, $\log \left(L_{\rm acc}/L_\sun\right) = 1.26 \log \left(L_{\rm Br\gamma}/L_\sun\right)+4.43$, the upper limit on the accretion luminosity would be $0.88~L_\sun$, far less than than the $\sim25~L_\sun$ estimated from the full SED.  A similar consideration applies to Pa$\beta$.  The mismatch between the accretion luminosity derived from \ion{H}{1} lines and the accretion luminosity derived from the SED is strong evidence for the FU Orionis nature of the V2775 Ori outburst as opposed to a lesser variety.

\subsection{Pinpointing the Breakdown of Magnetospheric Accretion}

At the onset of an FU Orionis outburst, the spectrum should change from that of a normal active accretor to that of an FU Orionis-like object.  This has been observed most famously in the spectrum of V1057 Cyg, which had optical hydrogen lines in emission in a pre-outburst spectrum obtained in 1957 \citep{her58} and the same lines in absorption in a spectrum obtained immediately after its 1969--1970 outburst \citep{her77}.  The same behavior has also been reported for V2493 Cyg (HBC 722 / LkH$\alpha$ 188-G4 / PTF 10qpf), which had H$\beta$ in emission in a 1977 spectrum \citep{coh79} and the same line in absorption just after its 2010 outburst \citep{mil11}.  Characterizing the range of accretion rates over which the spectrum makes this transition would be important in refining the theory of disk accretion in young stars.

While the physical basis for a tight correlation between the luminosity of Br$\gamma$ and the accretion luminosity in normally accreting young stellar objects is not completely understood, \citet{muz01} found that the fluxes and morphologies of near-IR hydrogen lines are reproduced well in models where the lines form in flows of matter from the accretion disk to the star along stellar magnetic field lines.  In general, the widths of the lines (${\rm FWHM}\sim200~{\rm km~s}^{-1}$) point to gas that free-falls from several stellar radii before landing on the star.  \citeauthor{muz01}\ found from their modeling that the correlations seem to be driven by the emitting area of the infalling gas:\ objects with larger accretion luminosities have larger emitting areas for the emission lines and thus larger line luminosities.

In this scenario, known as magnetospheric accretion, the accreting material free-falls to the star from the disk truncation radius $R_T$, the location where the pressure of the stellar magnetic field balances the ram pressure of the disk and the disk is truncated.  Magnetospheric accretion theory predicts $R_T\propto\dot{M}_{\rm disk}^{-2/7}$ if the stellar magnetic field, mass, and radius remain constant \citep{kon91}.  Thus if the accretion rate of a source engaged in magnetospheric accretion becomes several orders of magnitude larger, $R_T$ could shrink to the stellar radius.  For example, if $R_T=5~R_*$ when $\dot{M}_{\rm disk}\sim10^{-8}~M_\sun$~yr$^{-1}$ \citep{mey97}, then $R_T=R_*$ when $\dot{M}_{\rm disk}=2.8\times10^{-6}~M_\sun$~yr$^{-1}$.  This ``crushing of the magnetosphere'' would alter the line luminosity vs.\ accretion luminosity relationship that holds for magnetospheric accretion, producing much less \ion{H}{1} emission in spite of the dramatically larger accretion rate.  This is seen in the model H$\beta$ profiles of \citet{muz01}, which weaken and go into absorption for the largest accretion rates and the smallest flows.

The lack of Br$\gamma$ in the spectrum of V2775 Ori suggests that the model of accretion along magnetic field lines that thread the disk at $R_T>R_*$ does not apply for stars with accretion rates $\sim10^{-5}~M_\sun~{\rm yr}^{-1}$, an order of magnitude below the canonical accretion rate for FU Orionis objects.  Of the other young stellar objects with both documented outbursts and FU Orionis-like near-infrared spectra \citep{rei10,sem10,mil11,rei12}, most have luminosities of at least $100~L_\sun$.  V2493 Cyg is the only one with a post-outburst luminosity ($12~L_\sun$; \citealt{mil11}) similar to or lower than the $28~L_\sun$ found for V2775 Ori.  V2493 Cyg was a known classical T Tauri star prior to its outburst and has no protostellar envelope \citep{gre11}, making V2775 Ori the least luminous documented FU Orionis outburster with an envelope.  Both V2775 Ori and V2493 Cyg exist at the edge of the FU Orionis phenomenon and are important targets for understanding the transition between that regime and normal active accretion.

If V2775 Ori is beginning to fade as suggested by recent observations in the $K$ band, monitoring its spectrum as it fades may reveal a critical accretion luminosity at which \ion{H}{1} emission reappears, or it may show a gradual strengthening with decreasing accretion rate.  Continued observations of this source have the potential to probe magnetospheric accretion at its limit.

\section{CONCLUSIONS}

We have presented new near-IR spectra and imaging and far-IR-to-submillimeter photometry and imaging for V2775 Ori, an Orion protostar first identified to be in a state of enhanced luminosity by \citet{car11}.  By combining new and archival data, we conclude the following about the source:

\begin{enumerate}
\item The initial luminosity rise from 4.5 to $\sim51~L_\sun$, a factor of about 12, occurred between 2005 April 2 and 2007 March 12.  The total luminosity fell to $28.1~L_\sun$ by 2008 November 27.  Between 2009 February and 2011 January, the source dimmed slightly, by 0.25 mag at $K_s$.  The change in luminosity and the peak luminosity are at the low end of the ranges for previously known FU Orionis objects \citep{rei10}.
\item As a result of the outburst, the disk accretion rate of V2775 Ori increased from $\sim2\times10^{-6}~M_\sun~{\rm yr}^{-1}$ to $\sim10^{-5}~M_\sun~{\rm yr}^{-1}$, about an order of magnitude less than the canonical value for FU Orionis stars.  The envelope infall rate remained constant at about $7\times10^{-7}~M_\sun~{\rm yr}^{-1}$, less than the disk accretion rate at either epoch and indicating that the source is approaching the end of its envelope-dominated phase.
\item Our near-IR spectra of V2775 Ori are consistent with those of an object accreting mass at a rate toward the low end of the FU Orionis regime:\ they have CO absorption lines in the $K$ band with equivalent widths that are large for normal photospheres and small for FU Orionis stars, broad H$_2$O absorption, strong and wide blueshifted \helium\ absorption suggestive of a stellar or disk wind, and no atomic hydrogen emission, suggesting a lack of magnetospheric accretion.
\end{enumerate}

V2775 Ori has a luminosity, disk-to-star accretion rate, and near-IR spectrum consistent with an FU Orionis source, and it is the least luminous documented FU Orionis outburster with a protostellar envelope.  Additional near-infrared photometry and spectra of V2775 Ori will be instrumental in determining the time scale, if any, for the reappearance of its atomic hydrogen lines.  The existence of sources at the low-luminosity boundary of the FU Orionis regime supports the notion that there is a continuum of episodic accretion phenomena, not all of which result in sustained luminosities of order 100~$L_\sun$, which may explain some of the observed spread in protostellar luminosities.

\acknowledgments

Support for this work was provided by the National Aeronautics and Space Administration (NASA) through awards issued by the Jet Propulsion Laboratory, California Institute of Technology (JPL/Caltech).

This paper includes data from Herschel, a European Space Agency (ESA) space observatory with science instruments provided by European-led consortia and with important participation from NASA.  We also used the Spitzer Space Telescope and the Infrared Processing and Analysis Center (IPAC) Infrared Science Archive, which are operated by JPL/Caltech under a contract with NASA.  We include observations made under program 11548 of the NASA/ESA Hubble Space Telescope, obtained at the Space Telescope Science Institute, which is operated by the Association of Universities for Research in Astronomy, Inc., under NASA contract NAS 5-26555.

We include data from the Atacama Pathfinder Experiment, a collaboration between the Max-Planck-Institut f\"{u}r Radioastronomie, the European Southern Observatory, and the Onsala Space Observatory.  These data were collected at the European Southern Observatory, Chile, under proposal 088.C-0994.  We also made use of the Infrared Telescope Facility (IRTF), which is operated by the University of Hawaii under Cooperative Agreement NNX-08AE38A with NASA, Science Mission Directorate, Planetary Astronomy Program.  We reverently acknowledge the cultural significance of the Mauna Kea summit, site of the IRTF, to the indigenous Hawaiian community.

This paper makes use of data products from the Two Micron All Sky Survey, which is a joint project of the University of Massachusetts and IPAC/Caltech, funded by NASA and the National Science Foundation; the Wide-field Infrared Survey Explorer, which is a joint project of the University of California, Los Angeles, and JPL/Caltech, funded by NASA; and the Cornell Atlas of Spitzer/IRS Sources, which is a product of the Infrared Science Center at Cornell University, supported by NASA and JPL.

The work of A.~S.\ was supported by the Deutsche Forschungsgemeinschaft priority program 1573 (``Physics of the Interstellar Medium'').  M.~K.\ acknowledges support from the NSF-REU program of the Department of Physics and Astronomy at the University of Toledo (grant PHY-1004649).  M.~O.\ acknowledges support from MICINN (Spain) grants AYA2008-06189-C03 and AYA2011-30228-C03-01 (co-funded with FEDER funds).

Finally, we are grateful to Barbara Whitney and her collaborators for making their radiative transfer code available to the community and to Lynne Hillenbrand for helpful comments on a draft of this paper.

\end{document}